\documentclass[twoside,a4paper,11pt]{sea10}
\usepackage{graphicx}
\usepackage{hyperref}
\usepackage{movie15}
\usepackage{natbib}

\newcommand\DIB[1]{\mbox{DIB~$\lambda$#1}}
\newcommand\CaK{\mbox{Ca\,{\sc ii}~$\lambda$3934}}

\begin{document}
\pagenumbering{arabic}
\pagestyle{myheadings}
\thispagestyle{empty}
{\flushleft\includegraphics[width=\textwidth,bb=58 650 590 680]{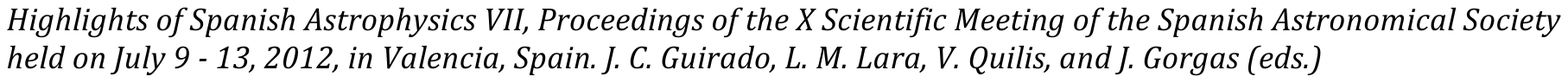}}
\vspace*{0.2cm}
\begin{flushleft}
{\bf {\LARGE
%
A study of the ISM with large massive-star optical spectroscopic surveys
%
}\\
\vspace*{1cm}
%
M Penad\'es Ordaz$^{1}$,
J. Ma\'{\i}z Apell\'aniz$^{1}$,
and
A. Sota$^{1}$ 
%
}\\
\vspace*{0.5cm}
%
$^{1}$
Instituto de Astrof\'{\i}sica de Andaluc\'{\i}a-CSIC, Glorieta de la Astronom\'{i}a s/n, \linebreak 18008 Granada, Spain
%
\end{flushleft}
%
\markboth{
A study of the ISM with large massive-star optical spectroscopic surveys
}{ 
%
Penad\'es Ordaz et al.
%
}
\thispagestyle{empty}
\vspace*{0.4cm}
\begin{minipage}[l]{0.09\textwidth}
\ 
\end{minipage}
\begin{minipage}[r]{0.9\textwidth}
\vspace{1cm}
\section*{Abstract}{\small
%
We are conducting a study on the imprint of the ISM on optical spectra based on two types of ongoing spectroscopic massive-star surveys: on 
the one hand, intermediate-resolution ($R = 2500$) green-blue spectra for $\sim$3000 stars obtained with the Galactic O Star Spectroscopic Survey 
(GOSSS). On the other hand, high-resolution ($R = 23\,000 - 65\,000$) optical spectra for 600 stars obtained from three different surveys,
OWN, IACOB, and NoMaDS. The $R = 2500$ data allows us to reach a larger sample with an average larger extinction while the 
$R = 23\,000 - 65\,000$ sample provides access to more diffuse interstellar bands (DIBs) and allows for the resolution in velocity of some ISM features. 
For each spectrum we are measuring the equivalent widths, FWHMs, and central wavelengths of 10-40 DIBs and interstellar lines (e.g. Ca~{\sc ii}~H+K, 
Na~{\sc i}~D1+D2) and, in the case of GOSSS, the existence of an H~{\sc ii} region around the star. We 
have also derived from auxiliary data or compiled from the literature values for the reddening, extinction law, H~{\sc i} column density, 
parallax, and H$\alpha$ emission. All of this constitutes the most complete collection ever of optical information on the ISM within 3 kpc of 
the Sun. We are analyzing the correlations between all of the collected quantities to discriminate between different possible origins of the 
DIBs. 
%
\normalsize}
\end{minipage}
%
%
\section{Introduction}

$\,\!$\indent The last two decades have seen an explosion of ISM studies based on the IR-to-radio wavelength ranges, in good part due to
large improvements in instrumentation and the number of dedicated telescopes at long wavelengths and the appearance of associated large-scale 
photometric and spectroscopic surveys. On the other hand, optical studies have mostly concentrated (with some notable exceptions such as
\citealt{Welsetal10}) in high-resolution analyses of small samples or
specific regions. Somehow, the large-scale optical surveys of absorption lines, pioneered e.g. by \citet{Adam49} or \citet{Duke51} and which
remained popular until the 1970s, have fallen out of fashion. This has happened despite the fact that some problems remain unsolved, such as the
nature of the diffuse interstellar bands (DIBs), whose carrier remains unclear almost a century after their discovery. Here we present a project
to do full-sky absorption-line surveys with larger samples, higher average extinctions, more modern instrumentation, and better S/N and complementary 
data than were possible
in the mid-twentieth century. Most of the stars in our sample are of O spectral type for two reasons: their spectra are cleaner than most other
stars (thus allowing to see the imprint of the ISM more easily) and their large luminosities allow us to probe to larger distances in the Galaxy.

\section{Data}

\subsection{GOSSS spectroscopy}

$\,\!$\indent The Galactic O-Star Spectroscopic Survey (GOSSS, \citealt{Maizetal11a}) is a 
project that is obtaining green-blue, intermediate resolution ($R\sim 2500$) spectra for $\sim 3000$ early-type stars with the immediate 
goal of spectrally classifying all of the bright Galactic O stars. As of late 2012, we have observed $\sim 1600$ stars. 
GOSSS data are being obtained with four different telescopes in both hemispheres: the 1.5 m telescope at the 
Observatorio de Sierra Nevada (OSN), the 3.5 m telescope at Calar Alto Observatory (CAHA), the 4.2 m William Herschel Telescope (WHT) at the
Observatorio del Roque de los Muchachos (ORM), and the 2.5 m du Pont telescope at Las Campanas Observatory (LCO). 
The first block of the survey was published as \citet{Sotaetal11a} and the second block will appear as Sota et al. (2013). 
GOSSS results on the spectral classification of O stars with C\,{\sc iii}~$\lambda\lambda$4647-4650-4652 in emission have appeared as 
\citet{Walbetal10a} and on rapidly rotating ON giants as \citet{Walbetal11}. 

GOSSS is a large-scale project with many goals, one of which is to study the imprint of the ISM on the observed spectra by measuring the 
absorption lines of atomic, molecular, and unknown (i.e. DIB) origin. Such studies have been traditionally carried out 
with high-resolution ($R > 10\,000$) echelle spectroscopy [1] to facilitate the detection of narrow, low equivalent width (EW) lines, [2] to 
study the Doppler and intrinsic profiles of the lines, and [3] to separate narrow ISM components from broad ones of stellar origin. However, 
intermediate-resolution spectra can also be used to study ISM lines of sufficient EW uncontaminated by stellar components. Furthermore, 
long-slit $R\sim2500$ data has three advantages over higher-resolution echelle data: [1] the ability to reach targets of fainter magnitudes 
(with the same S/N and exposure time), [2] an easier way of measuring broad absorption structures (with echelle data it can be hard to measure 
the position of the continuum if the absorption structure spans a range comparable to an echelle order), and [3] the posibility of placing two (or
more) targets on the slit, thus increasing the efficiency of the program.

\subsection{High-resolution spectroscopy}

$\,\!$\indent As discussed in the previous subsection, high-resolution spectroscopy is required to study some of the effects that the ISM
produces in the spectra of massive stars. For this project we are using four different high-resolution spectroscopic surveys of massive stars: 
OWN \citep{Barbetal10} is observing O and WN stars in the southern hemisphere using high-resolution spectrographs at La Silla, Las Campanas, 
Cerro Tololo (all of them in Chile), and CASLEO (Argentina). IACOB \citep{SimDetal11c,SimDetal11a} is observing northern O and B stars from La 
Palma (Spain). NoMaDS \citep{Maizetal11b} is extending the magnitude coverage of IACOB to fainter objects using the Hobby-Eberly Telescope at 
McDonald Observatory (USA). Finally, CAF\'E-BEANS has just started to follow up on IACOB and NoMaDS obtaining multiple epochs to detect spectrocospic 
binaries and calculate their orbits using the 2.2 m telescope at Calar Alto (Spain).
In total, we have observations of $\sim600$ stars with spectral resolutions between 23\,000 and 65\,000. In all cases
we reach down to 3900 \AA\ but the red limit varies between different setups from 7200 \AA\ to 10\,000 \AA. Some of the spectra have small gaps 
between orders or setups. The sample of stars observed at high resolution is nearly complete and has a lower mean extinction than the GOSSS
sample.

\subsection{Literature data}

$\,\!$\indent Here we list some of the sources for the complementary data used for this project:  

\begin{itemize}
  \item Spectral types are obtained from the GOSSS project itself \citep{Maizetal11a}.
  \item Photometry has been compiled from the Galactic O-Star Catalog, \href{http://gosc.iaa.es}{http://gosc.iaa.es} 
        \citep{Maizetal04b,Sotaetal08}.
  \item \citet{Gudeetal12} provides a modern compilation of interstellar column densities. 
  \item A number of sources \citep{JennDese94,Hobbetal08,Hobbetal09,Weseetal10} are used for existing information on DIBs.
  \item The new reduction of the Hipparcos data \citep{vanL07a} is used for the parallaxes.
\end{itemize}

\section{First results}

$\,\!$\indent We have started the study by measuring the EWs for 12 DIBs and three additional ISM lines for the O stars in our current GOSSS sample
using an interactive custom-made IDL procedure. Each individual measurement has been done first automatically and then inspected manually in order to tweak 
the positions of the fitted regions and 
background positions. This needs to be done because of the varying velocity shifts between the star and the intervening ISM, the different 
nearby stellar lines present for different spectral subtypes, and the need to establish a detection threshold as a function of S/N. In most cases the
EW has been obtained by numerical integration with a comparison value obtained by gaussian fitting in order to check for discrepancies. For some broad
DIBs (e.g. \DIB{4428}) it has been necessary to subtract from the profile overimposed stellar lines and in others (e.g. \DIB{4881}) the 
EW has been measured by fitting a Lorentzian profile due to the presence of a neraby strong contaminating stellar line. We have also checked that the 
measured central wavelength and FWHM yield consistent results. 

\begin{figure}
\centerline{
\includegraphics[width=\linewidth, bb=28 28 566 566]{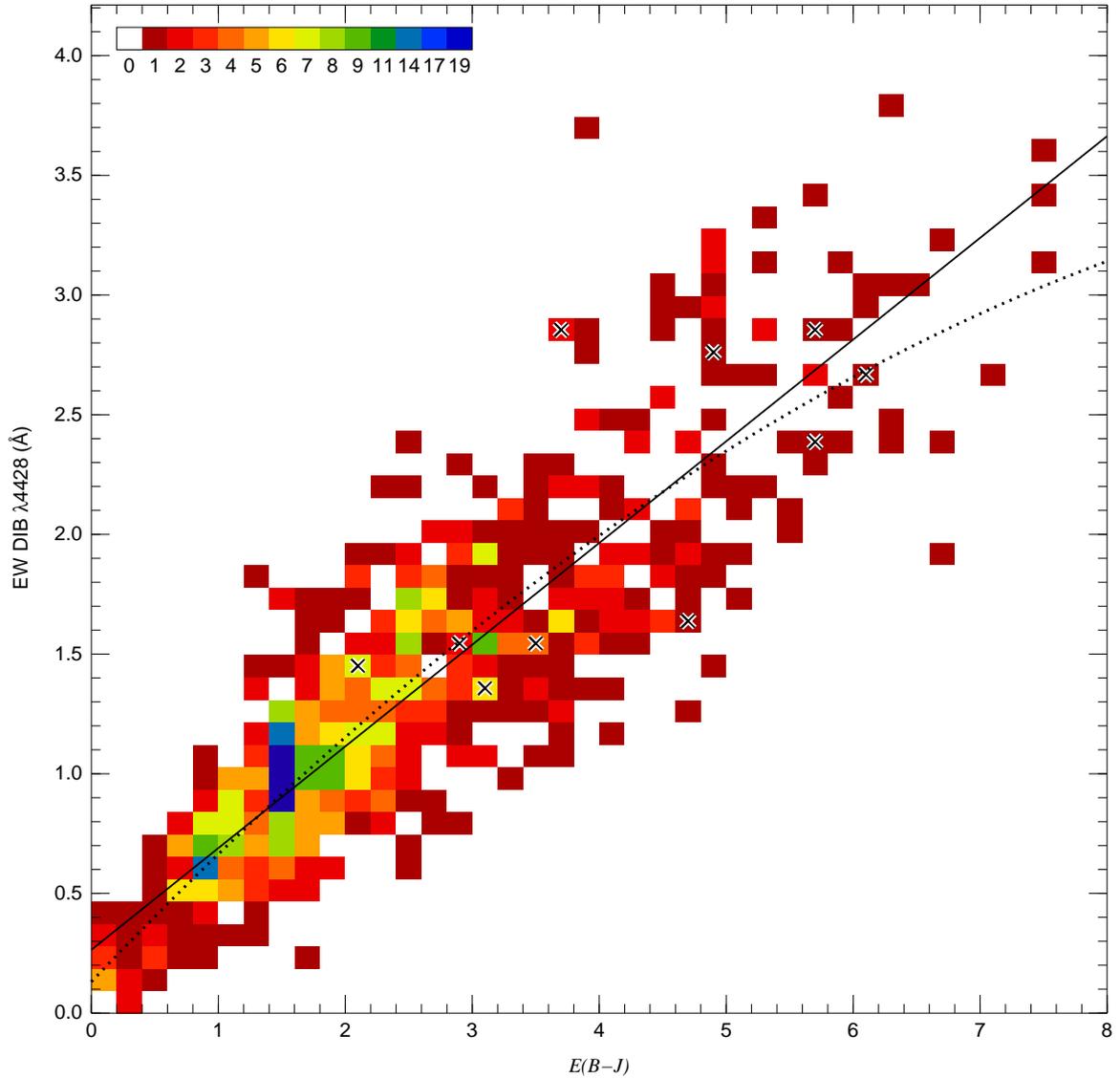}
}
\caption{2-D histogram of $E(B-J)$ against the EW of \DIB{4428}. The size of the vertical cell is set to the average uncertainty in the EW. Cells 
marked with an X have at least one point with large EW uncertainty. The continuous and dashed lines are first- and second-order polynomial fits, respectively.}
\end{figure}

\begin{figure}
\centerline{
\includegraphics[width=\linewidth, bb=28 28 566 566]{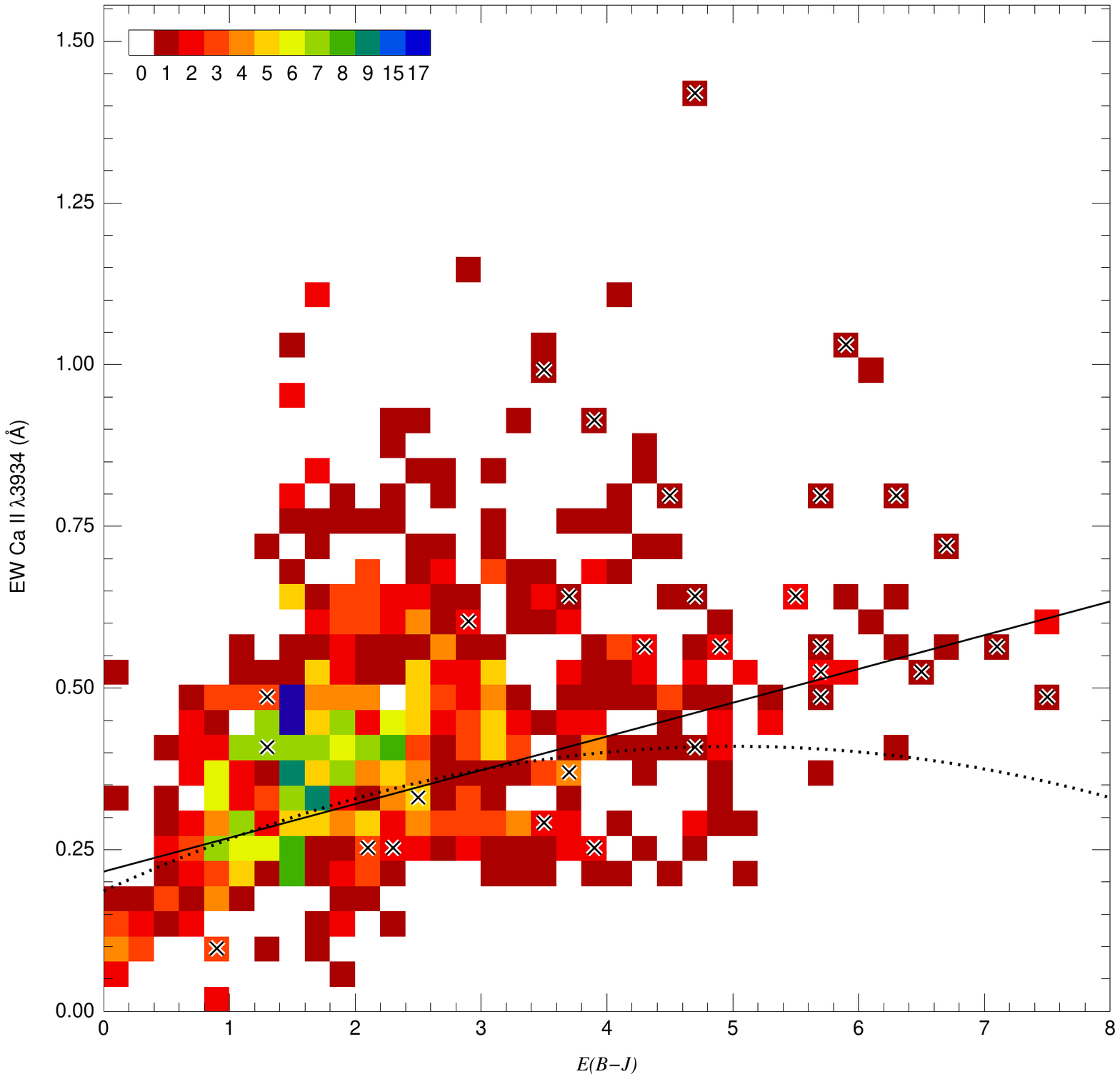}
}
\caption{Same as Fig.~1 for the EW of \CaK.}
\end{figure}

We show in Figs.~1~and~2 the 2-D histograms for the EWs of \DIB{4428} and \CaK\ with the color excess $E(B-J)$. As it can be seen, \DIB{4428} is highly (but not
completely) correlated with $E(B-J)$, a known result for DIBs and here extended to a larger, more extinguished (on average) sample. On the other hand, \CaK\ is
poorly correlated with $E(B-J)$: indeed, it is possible to find stars with twice the EW of other targets that are three or four times more extinguished. The reason
for the different behavior can be explained as follows: the DIB carriers are thought to be associated and well mixed with the diffuse ISM (e.g. \citealt{Herb93}), 
which is responsible for most of the extinction for our sample, and even the most intense \DIB{4428} is optically thin in our sample. Hence, \DIB{4428} is well 
correlated with $E(B-J)$. On the other hand, \CaK\ has a more uniform distribution than dust in the Galactic disk \citep{Megietal05a,Welsetal10}, leading to
more uniform column densities as a function of distance than dust. Also, \CaK\ saturates easily and this makes its EW strongly dependent on the kinematics of the
gas: a dispersion in cloud velocities helps increase the measured EW. Indeed, most points with EW \CaK\ $>$ 0.75 \AA\ in Fig.~2 are located in the Carina nebula, an 
object with extremely complex gas kinematics (typical lines of sight have 23 to 26 identified components, \citealt{Walbetal02c,Walbetal07}).

We have recently acquired new GOSSS spectra (not included in Figs.~1~and~2) that significantly increase the number of stars with $E(B-J) > 4$. Our goal is to publish
a paper during 2013 with the results derived from GOSSS spectroscopy. That paper will be followed by another one with the high-resolution data.

\section*{Acknowledgments}   
This research has made extensive use of [a] Aladin \citep{Bonnetal00} and [b] the SIMBAD database, operated at CDS, Strasbourg, France. 
Support for all authors was provided by [a] the Spanish Government Ministerio de Educaci\'on y Ciencia through grants AYA2010-17631 and AYA2010-15081
and [b] the Consejer{\'\i}a de Educaci{\'o}n of the Junta de Andaluc{\'\i}a through grant P08-TIC-4075. JMA also acknowledges support from
the George P. and Cynthia Woods Mitchell Institute for Fundamental Physics and Astronomy.

%

%
%
%
%
%
\bibliographystyle{aj}
\small
\bibliography{general}

\end{document}